\documentclass[prb,
twocolumn,
superscriptaddress,showpacs,amsmath,amssymb]{revtex4}
\usepackage{amsfonts}
\usepackage{bm}
\usepackage{verbatim}

\usepackage{graphicx}

\begin{document}
\title{From Standard Model of particle physics to room-temperature superconductivity}

\author{G.E.~Volovik}
\affiliation{Low Temperature Laboratory, Aalto University,  P.O. Box 15100, FI-00076 Aalto, Finland}
\affiliation{Landau Institute for Theoretical Physics, acad. Semyonov av., 1a, 142432,
Chernogolovka, Russia}

\date{\today}

\begin{abstract}
{Topological media are gapped or gapless fermionic systems, whose properties are protected by topology, and thus are robust to deformations of the parameters of the system and generic. We discuss here the class of gapless topological media, which contains the quantum vacuum of Standard Model in its symmetric phase, and also the condensed matter systems with zeroes in the fermionic energy spectrum, which form Fermi surfaces, Weyl and Dirac points, Dirac lines, Khodel-Shaginyan flat bands, etc. Some zeroes are  topologically protected, being characterized by topological invariants, expressed in terms of Green's function. For the stability of the others the ${\bf p}$-space topology must be accompanied by symmetry.

Vacua with Weyl points serve as a source of effective relativistic quantum fields
emerging at low energy: chiral fermions, effective gauge fields and tetrad gravity emerge together in the vicinity of a Weyl point. The accompanying effects, such as chiral anomaly,  electroweak baryo-production and chiral vortical effect,  are expressed via the symmetry protected ${\bf p}$-space invariants. 

The gapless topological media exhibit the bulk-surface and bulk-vortex correspondence: which in particular may lead to the flat band on the surface of the system or in the core of topological defects. The materials with flat band in bulk, on the surface or within the dislocations have singular density of states, which crucially influences the critical temperature of the superconducting transition in such media. While in all the known superconductors the transition temperature is exponentially suppressed as a function of the pairing interaction, in the flat band the transition temperature is proportional to the pairing interaction, and thus can be essentially higher. So the ${\bf p}$-space topology may give us the general recipe for the search or artificial fabrication of the room-temperature superconductors.
}
\end{abstract}

\maketitle

\section{Weyl point, level crossing and Berry phase monopole}
\label{LevelCrossing}

Some features of topological matter can be formulated in terms
of the Berry's connection in momentum space ${\bf p}$. 
For example, the Weyl points in the fermionic spectrum in Standard Model (SM) or in condensed matter Weyl materials such as superfluid $^3$He-A, can be viewed 
as the Berry's  phase monopoles  in momentum space  \cite{Volovik1987}.  Such monopole represents the topologically nontrivial point of level crossing  (the conical or diabolical point) in the 3D space of parameters, where the parameters are three components of the (quasi)particle momentum 
$(p_x,p_y,p_z)$.

\begin{figure}
\centerline{\includegraphics[width=1.1\linewidth]{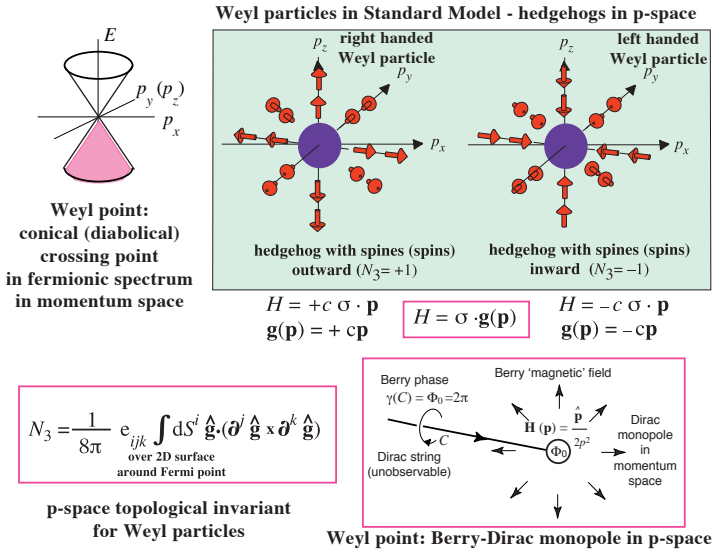}}
\caption{
\label{Fig:BerryMonopole}  
({\it top left}): The elementary crossing of levels in momentum ${\bf p}$ space, where only two branches of spectra touch each other.
Near the crossing the Hamiltonian is effectively reduced to $2\times 2$ matrix, expressed in terms of Pauli matrices and vector ${\bf g}({\bf p})$. 
({\it bottom left}): The crossing point
is characterized by topological invariant $N_3$. 
In the more general case of several species of interacting fermions
the topological charge $N_3$ is expressed in terms of the Green's function matrix in Eq.(\ref{MasslessTopInvariant3D}). 
({\it top right}): For $N_3=\pm 1$ the field of the 
 ${\bf g}({\bf p})$-vector has a hedgehog configuration in ${\bf p}$-space. Quasiparticles which live near the hedgehog with topological charge $N_3=+1$ 
behave as right-handed Weyl fermions of Standard Model, while those which live near the hedgehog with $N_3=-1$ 
behave as left-handed Weyl particles. Such crossing point in ${\bf p}$-space is called the Weyl point. ({\it bottom right}):  Weyl point represents the Berry phase Dirac monopole in ${\bf p}$-space.\cite{Volovik1987}
}
\end{figure}

Fig. \ref{Fig:BerryMonopole} demonstrates the elementary case of the level crossing, where only two branches of spectra touch each other.
Near the crossing point the complex $n\times n$ Hamiltonian is effectively reduced to the $2\times 2$ matrix
\begin{equation}
H= {\mbox{\boldmath$\sigma$}} \cdot {\bf g}({\bf p})\,,
\label{2x2}
\end{equation}
where ${\mbox{\boldmath$\sigma$}}$ are the Pauli matrices.
In the vicinity of the Weyl point ${\bf p}^{(0)}$, only linear terms in expansion of the Hamiltonian are important, and one obtains the effective Hamiltonian
\begin{equation}
H\approx  e^i_k\sigma^k(p_i- p^{(0)}_i)\,.
\label{2x2expansion}
\end{equation}
In the non-uniform situation the parameters of expansion $e^i_k$ and ${\bf p}^{(0)}$ become the fields, and then the Hamiltonian (\ref{2x2expansion}) describes the Weyl relativistic particle with spin $s=1/2$, which propagates in the 
effective gravitational field of the dreibein $e^i_k({\bf r})$ and in the effective electromagnetic field  ${\bf A}({\bf r})={\bf p}^{(0)}({\bf r})$.

This may suggest, that the Weyl particles of the Standard Model, the relativistic spin, 
the gauge and gravitational fields and all the other staff of relativistic quantum fields
could be the emergent phenomena, which originate from the level crossing in the underlying deep quantum vacuum \cite{Froggatt1991,Volovik2003}.
This is supported by the fact that the level crossing first discussed by von Neumann and Wigner \cite{NeumannWigner1929} is the robust phenomenon, which
is supported by topology \cite{Novikov1981}. That is why it may lead to the universal consequences which do not depend much on the microscopic details of the deep quantum vacuum.

In the simple case of $2\times 2$ complex matrix, the topological invariant can be expressed in terms of the unit vector $\hat{\bf g}={\bf g}/|{\bf g}|$, see Fig. 
\ref{Fig:BerryMonopole} ({\it bottom left}), where the integral is over the 2D surface around the Weyl point. This integer valued invariant $N_3$ determines the chiralty of the Weyl particle, which emerges near the crossing point. The fermions living near the Weyl point with topological charge $N_3=-1$ behave as the left-handed particles
with spectrum  $H=-c {\mbox{\boldmath$\sigma$}} \cdot {\bf p}$ , while
the value $N_3=+1$ corresponds to the right-handed fermions with $H=c {\mbox{\boldmath$\sigma$}} \cdot {\bf p}$. This suggests that the chiral nature of left-handed and right-handed quarks and leptons could be also the emergent phenomenon.
 
\section{Symmetry and topology of Green's functions}
\label{GreenFunctions}

The more interesting geometry and topology of the crossing points occur if the symmetries of the Hamiltonian are taken into account. For example, starting from purely real Hamiltonians $H$, one finds that depending on symmetry of $H$ the
effective quasiparticles emerging near the crossing point may behave as Majorana, Dirac or Weyl fermions. \cite{Horava2005,Wilczek2012,VolovikZubkov2014}
This suggests that appearance of complex numbers in quantum mechanics is also the emergent phenomenon, i.e. complex numbers emerge in the low energy description of the
underlying high energy theory together with Weyl particles and gauge and gravitational fields.

The symmetry consideration is especially important for Standard Model vacuum in its
symmetric semimetal phase above the electroweak transition. In this phase the total topological charge $N_3$ of the multiple crossing point at ${\bf p}=0$ is zero, $N_3=\sum_a  N_3^{(a)}=0$, where the sum is over all the fermions. The massless Weyl particles survive only due to a special symmetry between the fermions, which does not allow for mixing of Weyl points with opposite topological charges and thus protects the masslessness of quarks and leptons. 

However,  at this point the formulation in terms of the Berry phase becomes too complicated. Moreover, this formulation is not  complete, since it is applicable only to single-particle Hamiltonians and thus only to the non-interacting quantum fields. The more relevant formulation is in terms of the Green's function topology, since it automatically includes the interactions. In this approach the topological invariants (Chern numbers) can be expressed via the Green's function $G({\bf p},\omega)$, where $\omega$ is Matsubara frequency (see the earlier papers on the quantum Hall effect, quantum Hall topological insulators and Chern-Simons action
\cite{NiuThoulessWu1985,So1985,IshikawaMatsuyama1986,IshikawaMatsuyama1987,
Volovik1988,VolovikYakovenko1989,Yakovenko1989,Golterman1993}  and on topology of nodes in fermionic spectrum
\cite{GrinevichVolovik1988,Froggatt1991,Horava2005}, and the more recent papers \cite{EssinGurarie2011,GurarieEssin2013}).
  
The Weyl point represents the Green's function singularity described by the $\pi_3$ topological invariant: \cite{GrinevichVolovik1988}
 \begin{equation}
N_3 = \frac{e_{\alpha\beta\mu\nu}}{24\pi^2}~
{\bf tr}\int_\sigma   dS^\alpha
~ G\partial_{p_\beta} G^{-1}
G\partial_{p_\mu} G^{-1} G\partial_{p_\nu}  G^{-1}\,.
\label{MasslessTopInvariant3D}
\end{equation}
Here the integral is over the 3D surface $\sigma$ around the singular point   in the 4-momentum space   $p_\mu=(\omega,{\bf p})$.
For the non-interacting fermions one has $G^{-1}=i\omega -H$, and in case of $2\times 2$ matrix Hamiltonian the Eq.(\ref{MasslessTopInvariant3D}) transforms to the equation in  Fig. 
\ref{Fig:BerryMonopole} ({\it bottom left}) for the topological charge of the Berry phase monopole.

\begin{figure}
\centerline{\includegraphics[width=1.0\linewidth]{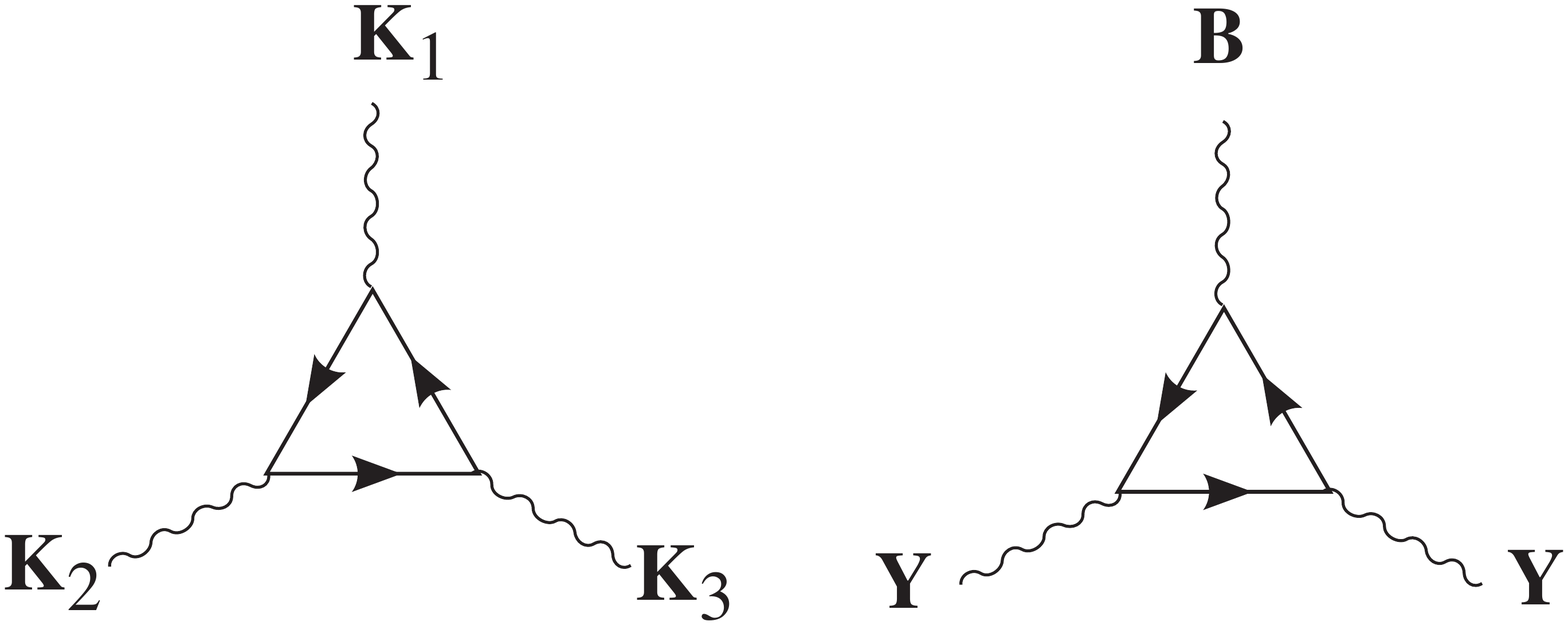}}
\caption{ \label{Fig:triangle}  Triangle
Feynman diagram, which is responsible for the phenomenon of chiral anomaly.
The wiggly lines correspond to the interaction of fermions with either gauge or effective fields.
In case of the baryoproduction caused by hyperelectric and hypermagnetic fields,
these are interactions with baryonic charge $B$ and hypercharges $Y$. 
The diagram on the right leads to Eq.(\ref{BarProdByHypercharge}) for the rate of the production
of the baryons from the vacuum. The prefactor in Eq.(\ref{BarProdByHypercharge}) is determined by the symmetry protected topological invariant $N_3({\cal K}_1,{\cal K}_2,{\cal K}_3)$
in Eq.(\ref{MultipleK}) with ${\cal K}_1={\cal B}$, the matrix of baryonic charges of fermions; and  ${\cal K}_2={\cal K}_3={\cal Y}$, the matrix of hypercharges. 
\\
In general, the ${\bf p}$-space topological invariant  $N_{2n+1}$ with $2n+1$ derivatives determines the prefactor in the Chern-Simons action in the $2n+1$ space-time dimension.\cite{Golterman1993,Volovik2003}
In particular, for $n=1$ the Chern-Simons action $S_{\rm CS}\propto N_3\int d^2xdtA d A$ describes the quantization of Hall conductivity in terms of topological charge $N_3$ in Eq.(\ref{MasslessTopInvariant3D}), where the integral is over $(\omega,p_x,p_y)$. 
Note, that the Chern-Simons terms in action should be treated with care, taking into account the bulk-boundary correspondence (see e.g. Ref. \cite{Chang2015}). 
}
\end{figure}

The Green's function approach also allows us to consider the combined effects of topology and symmetry, when the symmetry gives rise to additional topological invariants expressed via the Green's functions. \cite{Volovik1989,VolovikYakovenko1989}
 In case of the multiple Weyl points, the symmetry protected  topological invariants have the following structure: \cite{Volovik2000,Volovik2003}
 \begin{equation}
N_3({\cal K}) = \frac{e_{\alpha\beta\mu\nu}}{24\pi^2}~
{\bf tr}\left[{\cal K}\int_\sigma   dS^\alpha
~ G\partial_{p_\beta} G^{-1}
G\partial_{p_\mu} G^{-1} G\partial_{p_\nu}  G^{-1}\right]
\label{Ksymmetry}
\end{equation}
Here ${\cal K}$ is the matrix, which  commutes or anticommutes
with the Green's function matrix. This can be the discrete symmetry operator, the generator of continuous symmetry transformations, and in some cases the matrix of chemical potential
(see Eq.(20.6) for the chiral vortical efffect in Ref. \cite{Volovik2003}). 

In the symmetric vacuum of  Standard Model the key symmetry ${\cal K}$ is the $Z_2$ subgroup of the electroweak symmetry group $SU(2)\times U(1)$. \cite{Volovik2010}   It protects the
 masslessness of quarks and leptons, as a result the symmetric  phase of SM is analogous to Dirac  semimetals in condensed matter (identification of the topological Dirac semimetal in Cd$_3$As$_2$
has been reported in Ref. \cite{semimetal2014}). 
Below the electroweak transition the ${\cal K}$-symmetry is spontaneously broken. Without the symmetry protection, all the quarks and leptons acquire Dirac masses, and the vacuum of Standard Model becomes the topological insulator 
with its own topological invariants.\cite{Volovik2010}

\begin{figure}
\centerline{\includegraphics[width=1.0\linewidth]{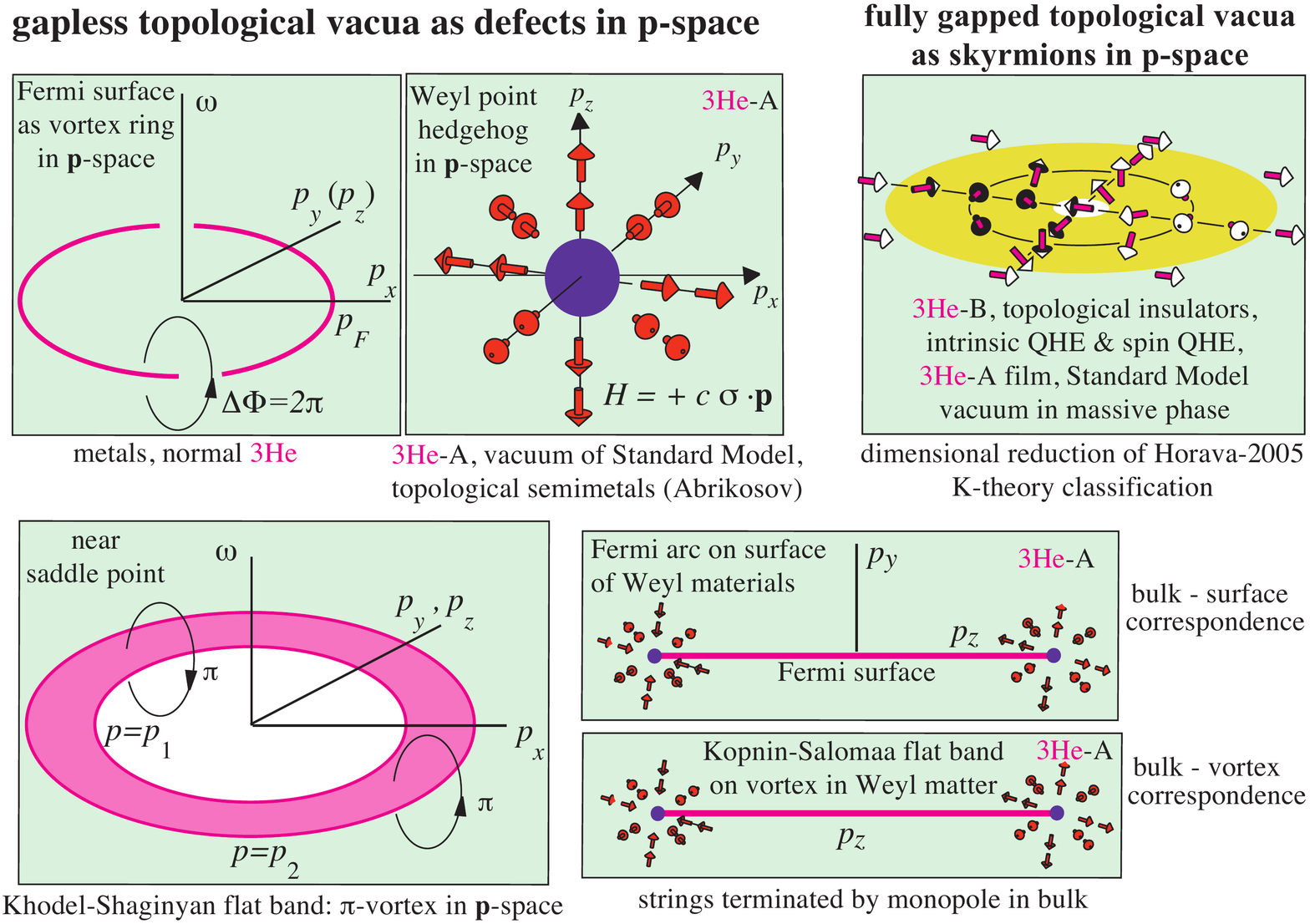}}
\caption{ 
\label{Fig:classes} 
 Topological materials as configurations in ${\bf p}$-space.  
\\
({\it top left}): Fermi surface as vortex in ${\bf p}$-space,\cite{Volovik2003} see Fig. \ref{Fig:FSVortexRing}. 
\\
({\it bottom left}): Khodel-Shaginyan flat band \cite{Khodel1990}, formed by splitting of Fermi surface (${\bf p}$-space vortex)
into two half-quantum vortices connected by ${\bf p}$-space domain wall.\cite{Volovik1991} 
\\
({\it top middle}): Weyl point as hedgehog (monopole) in ${\bf p}$-space (see Fig. \ref{Fig:BerryMonopole}). Topological stability of the Weyl point was first considered in Ref. \cite{NielsenNinomiya1981}. In Ref. \cite{VolovikMineev1982}  the Weyl point in $^3$He-A is viewed as the ${\bf p}$-space analog of boojum, the topological object introduced by David Mermin,\cite{Mermin1977} which can live on a surface of the system, but not in bulk  (topological classification of boojums see in Ref. \cite{Volovik1978}). In $^3$He-A the ${\bf p}$-space boojum lives on Fermi surface.
\\
({\it bottom right}):  Kopnin-Salomaa flat band in the vortex
core in the Weyl superfluids \cite{KopninSalomaa1991,Volovik1994,Volovik2011} 
and Fermi arc on the surface of the Weyl material \cite{Burkov2011} represent the ${\bf p}$-space analogs of strings terminated by monopole.
\\
 ({\it top right}): Topological insulators and fully gapped superfluds/superconductors are the nonsingular topological configurations in ${\bf p}$-space. Figure demonstrates the configuration of the field ${\bf g}(p_x,p_y)$ in the 2D topological insulator discussed in Ref. \cite{VolkovPankratov1985}.  This configuration is the ${\bf p}$-space analog of continuous topological objects --
skyrmions. The topological charge of this skyrmion, $N_3= \frac{1}{4\pi} ~
   \int    d^2p  
~\hat{\bf d}\cdot \left(\frac{\partial \hat{\bf d}}{\partial p_x}
\times \frac{\partial \hat{\bf d}}{\partial p_y} \right)$,  determines the intrinsic quantum Hall effect (i.e. without external magnetic field): 
$\sigma_{xy}= N_3e^2/h$.\cite{Volovik1988} In 2D crystals the integral is over the Brillouin zone.  
 The more general ${\bf p}$-space skyrmions describe the 3D topological band 
 insulators; \cite{HasanKane2010,Xiao-LiangQi2011} fully gapped superluid 
 $^3$He-B;\cite{SalomaaVolovik1988,Schnyder2008,Kitaev2009,Mizushima2014}  2D materials exhibiting intrinsic  quantum Hall and spin-Hall effects, such as gapped graphene,\cite{Haldane1988} thin film of superluid $^3$He-A and 2D planar phase of triplet 
 superfluid; \cite{Volovik1988,VolovikYakovenko1989,Yakovenko1989} chiral superconductor 
Sr$_2$RuO$_4$,\cite{Mackenzie2003} etc. These materials have the topological properties similar to that of  quantum vacuum of Standard Model in its broken symmetry massive phase, i.e. in the state below the electroweak transition, where
all elementary particles are massive.\cite{Volovik2010}
}
\end{figure}

The symmetry protected topological invariants 
\begin{eqnarray}
N_3({\cal K}_1,{\cal K}_2,{\cal K}_3) = 
\nonumber
\\
\frac{e_{\alpha\beta\mu\nu}}{24\pi^2}~
{\bf tr}\left[{\cal K}_1{\cal K}_2{\cal K}_3\int_\sigma   dS^\alpha
~ G\partial_{p_\beta} G^{-1}
G\partial_{p_\mu} G^{-1} G\partial_{p_\nu}  G^{-1}\right]
\label{MultipleK}
\end{eqnarray}
are responsible for quantization of different physical parameters. 
 For example, the production
of baryonic charge from the quantum vacuum by hyperelectric ${\bf E}_{Y}$ and
hypermagnetic ${\bf B}_{Y}$ fields is determined by the triangle diagram
in Fig. \ref{Fig:triangle}, which describes the phenomenon of 
chiral anomaly.\cite{Adler1969,BellJackiw1969}
The production rate is 
\begin{equation}
\dot B=\frac{1}{4\pi^2}N_3({\cal B} {\cal Y} {\cal Y}) ~{\bf B}_Y\cdot {\bf E}_{Y} \,,
\label{BarProdByHypercharge}
\end{equation}
where the coefficient $N_3({\cal B} {\cal Y} {\cal Y})$ is the topological invariant in Eq.(\ref{MultipleK}), with
${\cal K}_1$ being the matrix of baryonic charges $B$ of SM fermions; and  ${\cal K}_2={\cal K}_3={\cal Y}$,  the generator of the hypercharge $Y$.

Using ${\cal K}_1={\cal Q}$, the matrix of electric charges of SM fermions, and  
${\cal K}_2={\cal K}_3= {\cal \mu}$,  the matrix of chemical potentials, one obtains the prefactor in the so-called chiral vortical effect (CVE), when the dissipationless
equilibrium current is proportional to the angular velocity of rotation: 
${\bf J}=N_3({\cal Q} {\cal \mu} {\cal \mu}) ~{\bf \Omega}/h^2$.\cite{Volovik2003}
CVE has been studied in chiral liquids with the imbalance between the left and right fermions.
\cite{VilenkinLeahy1982,VolovikVilenkin2000,SonSurowka2009,Polikarpov2014}

\section{Fermi surface, flat band and room-temperature superconductivity}
\label{FS}

Fig. \ref{Fig:classes} demonstrates different classes of topological materials.
They correspond to different topological configurations in ${\bf p}$-space: quantized vortices (Fermi surfaces in metals); hedgehogs (Weyl materials); skyrmions (topological insulators and fully gapped topological superfluids);  half-quantum vortices terminating the domain wall (Khodel-Shaginyan flat band \cite{Khodel1990,Volovik1991}), the ${\bf r}$-space counterpart of this configuration can be found in Ref. \cite{Volovik1990}; and strings terminated by monopoles (Kopnin-Salomaa flat band in the core of Weyl superfluids \cite{KopninSalomaa1991,Volovik1994,Volovik2011}).

\begin{figure}
\centerline{\includegraphics[width=1.1\linewidth]{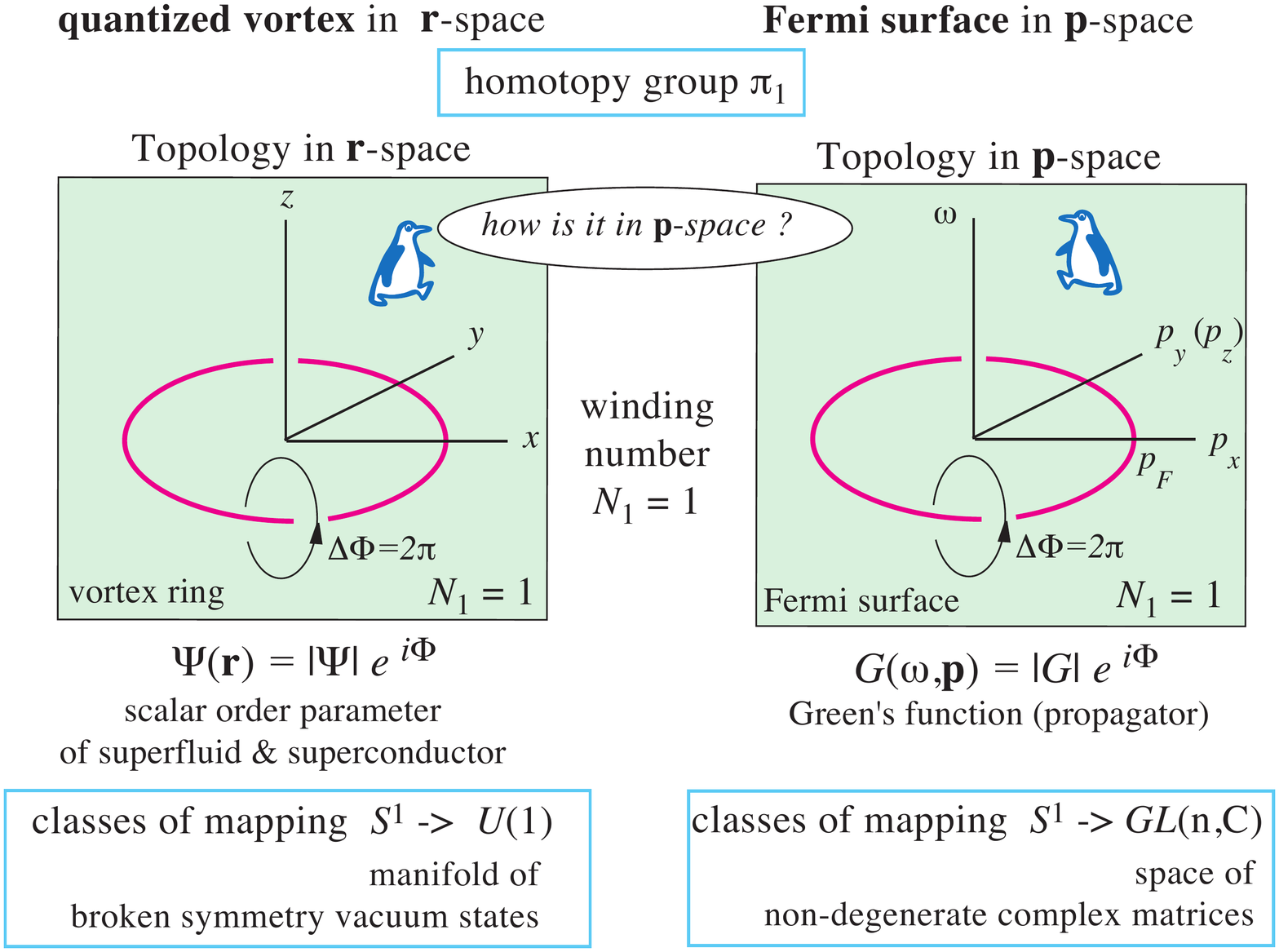}}
\caption{ \label{Fig:FSVortexRing}  ({\it left}): vortex ring in superfluid or superconductor. The phase $\Phi$ of the complex scalar order parameter changes by $2\pi$
around the vortex line (winding number $N_1=1$). ({\it right}): Fermi surface as quantized vortex line in
${\bf p}$-space. For simplicity one dimension ($p_z$) is suppressed, and the Fermi surface represents the closed line in 3D space $(\omega,p_x,p_y)$. In the simplest case, the Green's function is the complex scalar, and its phase $\Phi$ changes by 
$2\pi$
around the vortex line (winding number $N_1=1$). The nonzero winding number indicates, that the Green's function must have singularity on the line, which means 
the existence of the gapless fermionic excitations. In a general case, when the Green's function is matrix, the $N_1$ is the integer valued topological invariant, which describes the mapping of the contour around the vortex line to the space
of non-degenerate complex matrices.
}
\end{figure}

The most robust object in ${\bf p}$-space is the  Fermi surface. Its stability  is similar to the topological stability of the vortex line in superconductor
 (Fig. \ref{Fig:FSVortexRing}). Fermi surface  in ${\bf p}$-space  and the quantized vortex  in ${\bf r}$-space, though they live in different spaces, are described by the same topological invariant -- the winding number $N_1$, which characterizes the elements of homotopy group $\pi_1$. The topological stability of the Fermi surface towards perturbative interactions is at the origin of existence of metals.

\begin{figure}
\centerline{\includegraphics[width=0.8\linewidth]{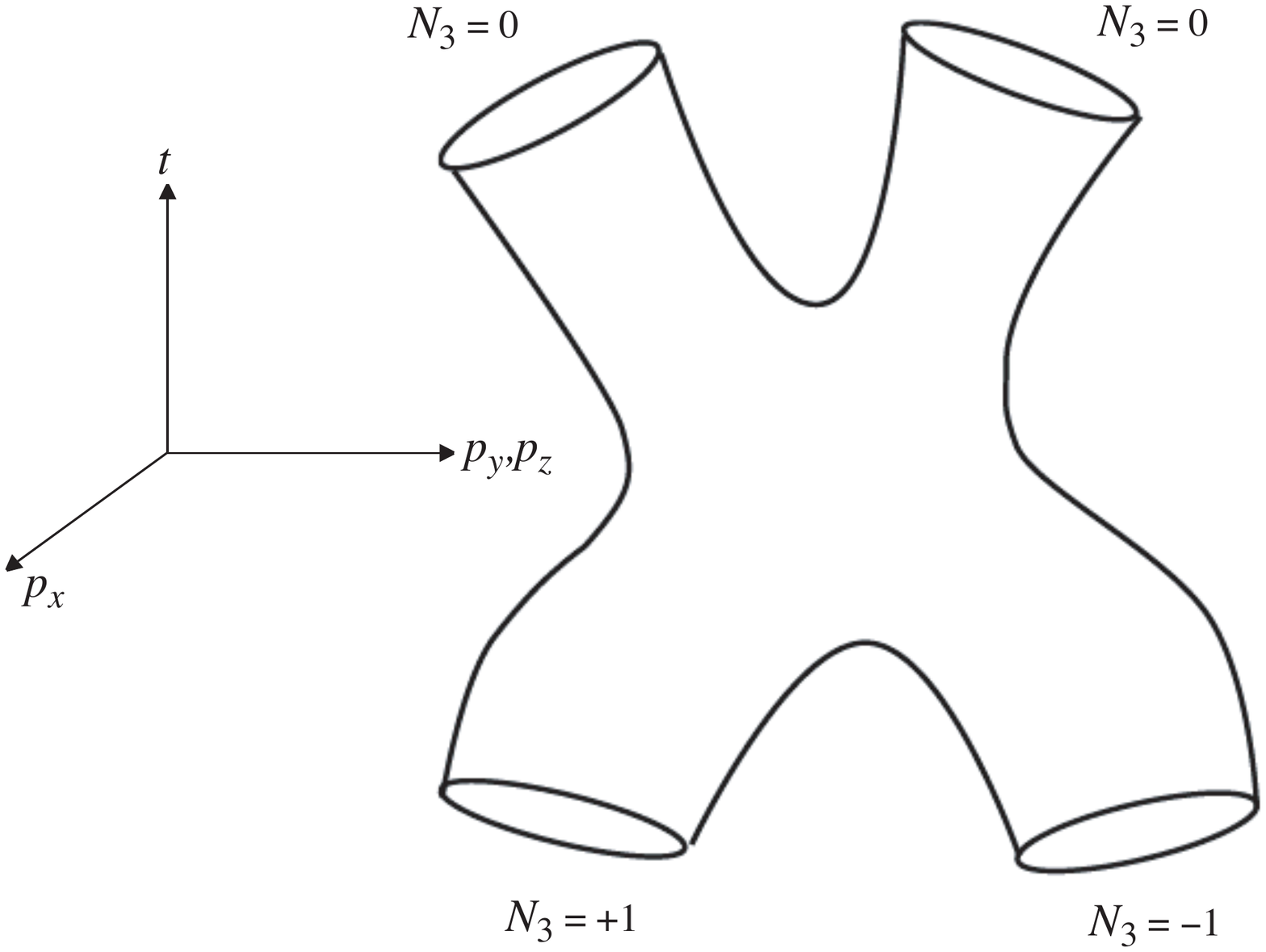}}
\caption{ \label{Fig:Collision2FS} Illustration of the interplay of local $N_1$ and global $N_3$ topological charges of Fermi surfaces.  The process of the topological quantum phase transtion is shown, where $t$ is the parameter of interaction, or time.  When $t$ increases, the initial Fermi spheres with opposite global  topological charges, $N_3=\pm 1$, merge  forming the single Fermi surface with $N_3=0$, which then
splits into two Fermi spheres, both with trivial global
 charges, $N_3=0$. \cite{KlinkhamerVolovik2005}
Looks somewhat similar to interaction of closed strings in string theory. 
The more interesting effects are expected if one takes into account the topology of the shape of the Fermi surface and also the other topological invariants, such as $N_2$, which describes the Dirac nodal lines in 3D systems and Dirac point nodes in graphene, \cite{HeikkilaVolovik2011,HeikkilaVolovik2015} see e.g.  Dirac lines within the Fermi surface of graphite.\cite{Mikitik2006,Mikitik2008} 
}
\end{figure}

The Dirac, Weyl and Majorana points, which are described by the homotopy group $\pi_3$, are rear in condensed matter systems. 
The reason for that is that the $\pi_3$ topological stability does not prevent the expansion of a point to a small Fermi surface. The topological invariant $N_3$ in Eq.(\ref{MasslessTopInvariant3D}) still holds if the integral is around the whole Fermi pocket. 
So such Fermi surface
has two topological invariants, the local invariant $N_1$ of a ${\bf p}$-space vortex and the global invariant $N_3$
of the former ${\bf p}$-space hedgehog. An example of the interplay of two invariants, $N_1$ and $N_3$, is demonstrated in Fig. \ref{Fig:Collision2FS}. This figure describes the  topological quantum phase transitions at which the global invariant $N_3$ is transferred between two Fermi spheres. \cite{KlinkhamerVolovik2005}
 In addition, the topological invariant $N_2$, which describes the Dirac nodal lines,\cite{HeikkilaVolovik2015} can be involved  leading to a more complicated behavior, see e.g. the structure of four Dirac lines which are expanded 
to form the Fermi surface of graphite.\cite{Mikitik2006,Mikitik2008} 
 
\begin{figure}
\centerline{\includegraphics[width=1.1\linewidth]{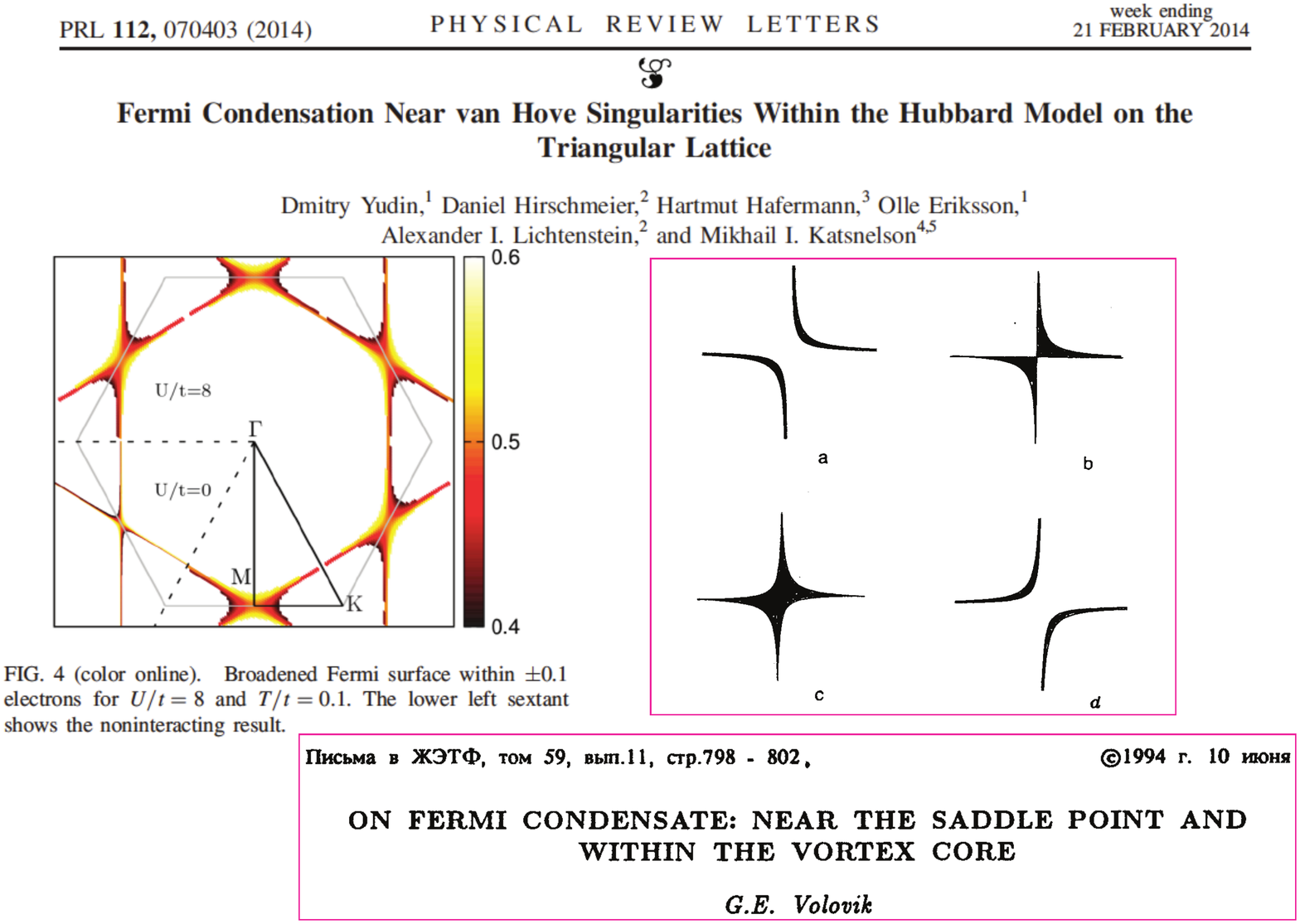}}
\caption{ \label{Fig:FermiCondensate}  Formation of the Khodel-Shaginyan
flat band (Fermi condensate) near the saddle point caused by interaction between electrons.
The right Figure demonstrates the results following from the
phenomenological Landau model of Fermi liquid.\cite{Volovik1994} As the parameter of interaction changes the Lifshitz transition (the reconnection of the Fermi surfaces) splits into the chain of the quantum phase 
transitions with the flat bands in  intermediate states. (a) two regions of the flat band are formed; (b) the two flat bands merge to form a single flat band in (c);
(d) flat band splits in two flat bands, which finally disappear.  
The left Figure shows results of the numerical simulations of the Hubbard model.\cite{Yudin2014}
The strong interaction leads to the flattening of the spectrum in the vicinity of the
saddle point.
}
\end{figure}

If the interaction  between the electron is strong enough, the Fermi surface in turn can be expanded to the more powerful manifold of zeroes, where the energy of electrons is zero in the 3D region  of momentum space \cite{Khodel1990,Volovik1991}. Such 3D flat band is formed due to the effect of merging of the energy  levels caused by the interaction between the fermions.  Both the phenomenological Landau type theory \cite{Volovik1994}  and the Hubbard model calculations \cite{Yudin2014} suggest, that the interaction supported flat band is more easily formed in the vicinity of the saddle point (see Fig. \ref{Fig:FermiCondensate}).
Recently the merging of levels due to interaction has been reported in Ref. \cite{Dolgopolov2014}.
From the point of view of the ${\bf p}$-space topology, the flat band is formed by splitting of the 
${\bf p}$-space vortex to two half-quantum vortices terminating the ${\bf p}$-space domain wall -- the flat band,\cite{Volovik1991} see 
Fig. \ref{Fig:classes} ({\it bottom left}).

\begin{figure}
\centerline{\includegraphics[width=1.2\linewidth]{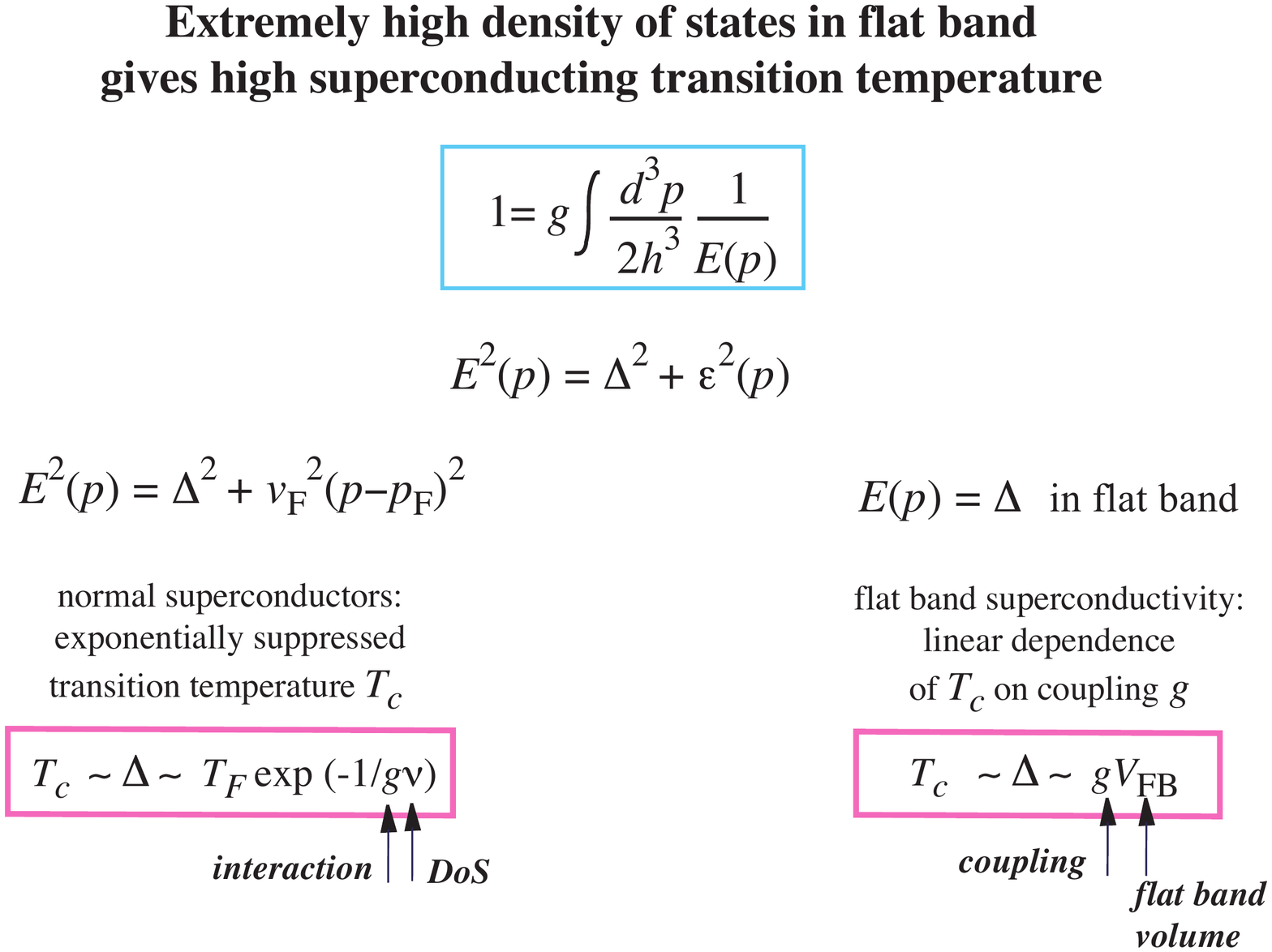}}
\caption{ \label{Fig:roomT}  The gap equation ({\it top}) leads to exponentially suppressed transition temperature in conventional Fermi liquid  ({\it left}) and to the linear dependence of the superconducting transition temperature
on the coupling $g$ in the flat band material  ({\it right}).\cite{Khodel1990,KopninHeikkilaVolovik2011} The flat band may open the route to room temperature
superconductivity.
}
\end{figure}

The flat band has the enhanced density of states, as compared to the Fermi surface. As a result the superconducting transition temperature is not exponentially suppressed, but is proportional to the coupling constant $g$ in the Cooper channel
\cite{Khodel1990} (see Fig. \ref{Fig:roomT}). The flat band materials have a good perspective for achievement of room temperature superconductivity. 

\section{Topological flat bands: another route to room-T superconductivity}
\label{TopFlat}

\begin{figure}
\centerline{\includegraphics[width=1.1\linewidth]{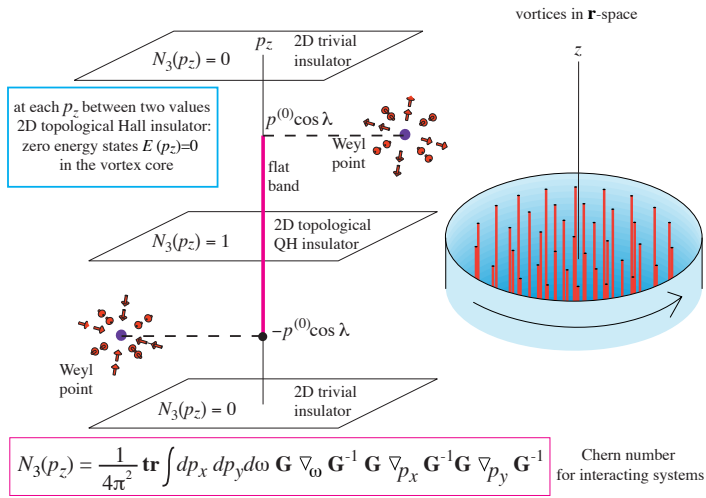}}
\caption{ \label{Fig:CoreFlatBand} Topological origin of the Kopnin-Salomaa 
flat band of Majorana fermions in the core of a vortex in Weyl superflud/superconductor \cite{KopninSalomaa1991}.
The vortex is oriented along the axis $z$. The angle $\lambda$ is the polar angle of the
position ${\bf p}^{(0)}$ of the Weyl point in ${\bf p}$-space. The planes $p_z={\rm const}$ represent the 2D 
superfluids. For $|p_z| \neq  p^{(0)} |\cos\lambda|$, the energy spectrum $E_{p_z}(p_x,p_y)$
is fully gapped. For $|p_z| < p^{(0)} |\cos\lambda|$ the 2D superfluids are topological, since their
topological invariant on the bottom of the Figure is $N_3(p_z)=1$. The vortex core in the topological 2D superfluid
contains the Andreev-Majorana bound state with zero energy \cite{Volovik1999,ReadGreen2000}. The set of the Andreev-Majorana bound states
form the flat band in the interval $|p_z| < p^{(0)} |\cos\lambda|$.
The flat band terminates at points where the bulk spectrum has zero, that is why
it represents the string in the ${\bf p}$-space, which is terminated by monopole in bulk
in Fig.\ref{Fig:classes} ({\it bottom right}).
}
\end{figure}

The momentum space topology provides the other sources of the flat band, which do not depend on interaction. The flat band emerges due to the correspondence between the topology 
in bulk and the topology of spectrum on the surface (bulk-surface correspondence), or within the topological object (bulk-vortex correspondence). Let us consider the
formation of the topological flat band on example of the Kopnin-Salomaa flat band of Andreev-Majorana fermions
in the vortex core of Weyl superfluids \cite{KopninSalomaa1991} (see Fig. \ref{Fig:CoreFlatBand} and Ref. \cite{Volovik2011}). 

If the vortex is oriented along  the  $z$-axis, then $p_z$ is a good quantum number both for  the bulk states and the bound states in the vortex core. So one can consider the problem at fixed value of $p_z={\rm const}$ as a parameter. In other words we have a set of 2D fermionic systems numerated by the parameter $p_z$, with the bulk spectrum $E_{p_z}(p_x,p_y)$. 
 
If the plane $p_z={\rm const}$ does not contain the Weyl point, at which
$E_{p_z}(p_x,p_y)=0$, this plane represents the fully gapped 2D superfluid. Such superfluid is characterized by the topological invariant $N_3$ in Fig. \ref{Fig:CoreFlatBand} ({\it bottom}). \cite{So1985,IshikawaMatsuyama1986,IshikawaMatsuyama1987,Volovik1988,VolovikYakovenko1989}
 The value of this invariant depends on $p_z$.
For $|p_z| < p^{(0)} |\cos\lambda|$ one has $N_3(p_z)=1$, which means that for these $p_z$ the 2D superfluids are topological, while for $|p_z| > p^{(0)} |\cos\lambda|$ the 2D superfluids are topologically trivial with $N_3(p_z)=0$.
Point vortex in the topologically nontrivial chiral 2D superfluids 
has the Majorana mode with exactly zero energy.\cite{Volovik1999,ReadGreen2000} That is why the vortex line in the original 3D system has the dispersionless branch of Majorana modes, $E_{\rm bound}(p_z)=0$, for all $p_z$ in the interval $|p_z| < p^{(0)} |\cos\lambda|$. This is the flat band.

The flat band terminates at points where it merges with the bulk spectrum $E_{p_z}(p_x,p_y)$,
i.e. the boundary of the flat band are determined by the positions ${\bf p}^{(0)}$ of the Weyl point in ${\bf p}$-space.

In a similar manner the flat band appears on the surface of the 3D system, if the bulk contains topologically
protected lines of zeroes (Dirac lines).\cite{Ryu2002,SchnyderRyu2011,HeikkilaVolovik2011,HeikkilaKopninVolovik2011,SilaevVolovik2014,HeikkilaVolovik2015,Kane2015}
According to the bulk-surface correspondence, the boundaries of the flat band are determined by the projection of the Dirac line to the surface. In the lower dimension the same effect leads
to the 1D flat band  on the zig-zag boundary of 2D graphene.\cite{Ryu2002}

Recently another possible source of the topological flat band has been discussed in two materials:
 highly oriented pyrolytic Bernal  graphite (HOPG) \cite{EsquinaziHeikkila2014} and heterostructures SnTe/PbTe, PbTe/PbS, PbTe/PbSe, and PbTe/YbS consisting of a topological crystalline insulator and a trivial insulator.\cite{TangFu2014} In both cases the flat band
comes from the misfit dislocation array, which is spontaneously formed at the interface between two  crystals due to the lattice mismatch.  In Ref. \cite{EsquinaziHeikkila2014} the lattice of screw dislocations has been considered, which emerges at the interface between two domains of HOPG with different orientations of crystal axes. In Ref. \cite{TangFu2014} the misfit dislocation array is formed at the interface between topological and trivial insulators. The topological origin of the flat bands in these systems can be understood in terms of the overlapping of the 1D flat bands formed within the dislocations. 

The above two systems exhibit similar phenomenon. In both cases the surface superconductivity is reported,
which is concentrated at the interfaces. The reported transition temperature essentially exceeds  the typical transition temperature expected for the bulk materials. The possible origin of this pehnomenon is the flat band at the interfaces, where the transition temperature could be proportional to the coupling constant and the area of the flat band (see Fig. \ref{Fig:roomT}).
This suggests to seriously reconsider different publications reporting
superconducting-like signals up to room temperature 
in graphite-based samples (see Ref. \cite{Ballestar HeikkilaEsquinazi2014} and references therein).


\begin{thebibliography}{99}

 
\bibitem{Volovik1987}
G.E. Volovik,  
Zeroes in the fermionic spectrum in superfluid systems as diabolical points,
JETP Letters 46, 98 (1987).

\bibitem{Froggatt1991}
C.D. Froggatt   and  H.B. Nielsen,
{\it Origin of Symmetry}, 
World Scientific, Singapore, 1991.

\bibitem{Volovik2003} 
G.E. Volovik, 
{\it The Universe in a Helium Droplet}, 
Clarendon Press,  Oxford (2003).

\bibitem{NeumannWigner1929} 
J. von Neumann J and E. Wigner,
Phys. Z. {\bf 30}, 467 (1929).

\bibitem{Novikov1981}
S.P. Novikov, 
 Magnetic Bloch functions and vector bundles. Typical dispersion laws and their quantum numbers, 
 Sov. Math., Dokl. {\bf 23}, 298--303 (1981).

\bibitem{Horava2005}  
P. Ho\v{r}ava,
Stability of Fermi surfaces and $K$-theory,
Phys. Rev. Lett. \textbf{95}, 016405 (2005).

\bibitem{Wilczek2012}
F. Wilczek,
Introduction to quantum matter,
Phys. Scr. T{\bf 146},  014001 (2012).

\bibitem{VolovikZubkov2014}
G.E. Volovik and M.A. Zubkov,
Emergent Weyl spinors in multi-fermion systems,
Nuclear Physics B {\bf 881}, 514--538  (2014);
arXiv:1402.5700.

\bibitem{NiuThoulessWu1985} 
Qian Niu, D. J. Thouless, and Yong-Shi Wu,
Quantized Hall conductance as a topological invariant,
Phys. Rev. {\bf B~31}, 3372--3377 (1985).

\bibitem{So1985} 
H. So,
Induced topological invariants by lattice fermions in odd dimensions,
Prog. Theor. Phys. {\bf 74}, 585--593 (1985).

\bibitem{IshikawaMatsuyama1986} 
K. Ishikawa  and T. Matsuyama,
Magnetic field induced multi component QED in three-dimensions and quantum Hall effect,
Z. Phys. C {\bf 33}, 41--45 (1986). 

\bibitem{IshikawaMatsuyama1987} 
K. Ishikawa and T. Matsuyama,
A microscopic theory of the quantum Hall effect, 
Nucl. Phys. {\bf B~280}, 523--548  (1987).

\bibitem{Volovik1988}
 G.E. Volovik, 
 An analog of the quantum Hall effect in a superfluid 3He film,
JETP  {\bf 67}, 1804 (1988).
 
\bibitem{VolovikYakovenko1989}  
G.E. Volovik   and V.M. Yakovenko,  
Fractional charge, spin and statistics of solitons in superfluid $^3$He film, 
J. Phys.: Condens. Matter {\bf 1},  5263--5274 (1989).

 \bibitem{Yakovenko1989} 
V.M. Yakovenko,
Spin, statistics and charge of solitons in (2+1)-dimensional theories,   
Fizika (Zagreb) {\bf 21}, suppl. 3, 231 (1989); 
arXiv:cond-mat/9703195.

\bibitem{Golterman1993}
 M.F.L. Golterman, K.  Jansen and D.B. Kaplan,
Chern-Simons  currents and chiral  fermions on the lattice,
 Phys.Lett. B {\bf 301}, 219--223 (1993):
arXiv: hep-lat/9209003.

\bibitem{GrinevichVolovik1988} 
 P.G. Grinevich, G.E. Volovik,
 Topology of gap nodes in superfluid  3He:  $\pi_4$ homotopy group for 3He-B  disclination, 
J. Low Temp. Phys. {\bf 72}, 371  (1988).

 

\bibitem{EssinGurarie2011}
A.M. Essin, V. Gurarie,
Bulk-boundary correspondence of topological insulators from their Green's functions,
Phys. Rev. B {\bf 84}, 125132 (2011).

\bibitem{GurarieEssin2013} 
V. Gurarie, A.M. Essin,
Topological invariants for the fractional quantum Hall states,
 JETP Lett. {\bf 97}, 233-238 (2013);
arXiv:1301.3941.

\bibitem{Volovik1989}
 G.E. Volovik, 
 Fractional statistics and analogs of quantum Hall effect in superfluid $^3$He films, 
 AIP Conference Proceedings  {\bf 194}, 136--146 (1989).

 \bibitem{Volovik2000}
G.E. Volovik, 
Momentum-space topology of Standard Model, 
J. Low Temp. Phys.,  {\bf 119}, 241 -- 247 (2000); hep-ph/9907456.

 \bibitem{Volovik2010} 
 G.E. Volovik, 
Topological invariants  for Standard Model: from semi-metal to topological insulator,
JETP Lett. {\bf 91}, 55--61 (2010);
arXiv:0912.0502.

\bibitem{semimetal2014} 
Z. K. Liu,	 J. Jiang,	 B. Zhou,	 Z. J. Wang,	 Y. Zhang,	 H. M. Weng,	 
D. Prabhakaran,	 S-K. Mo,	 H. Peng,	 P. Dudin,	 T. Kim,	 M. Hoesch,	
 Z. Fang,	 X. Dai,	 Z. X. Shen,	 D. L. Feng,	 Z. Hussain	and  Y. L. Chen,
A stable three-dimensional topological Dirac semimetal Cd$_3$As$_2$,
Nature Materials {\bf 13}, 677--681 (2014).

\bibitem{Chang2015} 
Ming-Che Chang  and Min-Fong Yang,
Chiral magnetic effect in a two-band lattice model of Weyl semimetal,
Phys. Rev. B {\bf 91}, 115203 (2015).

\bibitem{Adler1969} 
S. Adler,  
Axial-vector vertex in spinor electrodynamics,
Phys. Rev. {\bf 177}, 2426--2438 (1969).  

\bibitem{BellJackiw1969}  
J.S. Bell   and R. Jackiw,
A PCAC puzzle: $\pi_0\rightarrow\gamma\gamma$ in the $\sigma$ model,
Nuovo Cim. A {\bf 60},  47--61 (1969).

\bibitem{VilenkinLeahy1982} 
A. Vilenkin  and D.A. Leahy,
Parity non-conservation and the origin of cosmic magnetic fields, 
Astrophys. J. {\bf 254}, 77--81  (1982).


\bibitem{VolovikVilenkin2000}  
G. E. Volovik and A. Vilenkin,
Macroscopic parity violating effects and $^3$He-A,
Phys. Rev. D {\bf 62}, 025014 (2000).

\bibitem{SonSurowka2009}  
D.T. Son and P.  Surowka,
Hydrodynamics with triangle anomalies,
Phys. Rev. Lett. {\bf 103}, 191601 (2009).

\bibitem{Polikarpov2014}
V. Braguta, M. N. Chernodub, V. A. Goy, K. Landsteiner, A. V. Molochkov, M. I. Polikarpov,
Phys. Rev. D {\bf 89}, 074510 (2014).


 \bibitem{NielsenNinomiya1981} 
H.B. Nielsen, M. Ninomiya: 
Absence of neutrinos on a lattice.  I - Proof by homotopy theory, 
Nucl. Phys. B \textbf{185}, 20  (1981); 
Absence of neutrinos on a lattice. II - Intuitive homotopy proof,  
Nucl. Phys. B \textbf{193}, 173 (1981). 

\bibitem{VolovikMineev1982} 
G.E. Volovik, V.P. Mineev, 
Current in  superfluid Fermi liquids and the vortex core structure,
JETP {\bf 56}, 579--586 (1982).

\bibitem{Mermin1977} 
 N.D. Mermin,
Surface singularities and superflow in $^3$He-A', 
in: {\it Quantum Fluids and Solids}, eds.
S. B. Trickey, E. D. Adams and J. W. Dufty, Plenum, New York, pp. 3--22  (1977).

\bibitem{Volovik1978} 
G.E. Volovik, 
Topological singularities on the surface of an ordered system,
JETP Lett. {\bf 28}, 59--62 (1978).

\bibitem{Khodel1990}
V.A. Khodel  and  V.R. Shaginyan,
Superfluidity in system with fermion condensate,
JETP Lett. \textbf{51}, 553 (1990).

\bibitem{Volovik1991}
G.E. Volovik, 
A new class of normal Fermi liquids,
{\it JETP Lett.} \textbf{53}, 222 (1991).

\bibitem{KopninSalomaa1991}
N.B.  Kopnin and M.M. Salomaa, 
Mutual friction in superfluid $^3$He: Effects of bound states in the vortex core,
Phys. Rev. B {\bf 44}, 9667--9677 (1991).

\bibitem{Volovik1994}  
G.E. Volovik, 
On Fermi condensate: near the saddle point and within the vortex core, 
JETP Lett. {\bf  59}, 830--835 (1994).

\bibitem{Volovik2011}  
G.E. Volovik,
Flat band in the core of topological defects: bulk-vortex correspondence in topological superfluids with Fermi points,
 JETP Lett. {\bf 93}, 66--69 (2011);
arXiv:1011.4665.

\bibitem{Burkov2011}
A.A. Burkov and L. Balents, 
Weyl semimetal in a topological insulator multilayer,
Phys. Rev. Lett. {\bf 107}, 127205 (2011);
A.A. Burkov, M.D. Hook, L. Balents,
Topological nodal semimetals,
Phys. Rev. B {\bf 84}, 235126 (2011).

\bibitem{VolkovPankratov1985}  
B.A. Volkov and O.A. Pankratov,
Two-dimensional massless electrons in an inverted contact,
JETP Lett. {\bf 42}, 178--181 (1985).

\bibitem{HasanKane2010}   
M.Z. Hasan and C.L. Kane, 
Topological Insulators,
Rev. Mod. Phys. {\bf 82}, 3045--3067 (2010).

\bibitem{Xiao-LiangQi2011}   
Xiao-Liang Qi and Shou-Cheng Zhang, 
Topological insulators and superconductors,
Rev. Mod. Phys. {\bf 83}, 1057--1110 (2011).

\bibitem{SalomaaVolovik1988}
 M.M. Salomaa and  G.E. Volovik, 
 Cosmiclike domain walls in superfluid $^3$He-B: Instantons and diabolical points in (${\bf k}$,${\bf r}$) space, Phys. Rev.  {\bf B~37}, 9298--9311 (1988).

\bibitem{Schnyder2008} 
A.P. Schnyder, S. Ryu, A. Furusaki and A.W.W. Ludwig, 
Classification of topological insulators and superconductors in three spatial dimensions,
Phys. Rev. {\bf B~ 78}, 195125 (2008); A.P. Schnyder, S. Ryu, A. Furusaki and A.W.W. Ludwig, 
Classification of topological insulators and superconductors,
 AIP Conf. Proc. {\bf 1134}, 10 (2009);    
 arXiv:0905.2029.

\bibitem{Kitaev2009} 
A. Kitaev,
Periodic table for topological insulators and superconductors,
AIP Conference Proceedings, Volume {\bf 1134}, pp. 22--30 (2009);
  arXiv:0901.2686.
  
  
  \bibitem{Mizushima2014}
 T. Mizushima, Y. Tsutsumi, M. Sato, K. Machida,
Symmetry protected topological superfluid $^3$He-B,
J. Phys.: Condens. Matter {\bf 27}, 113203 (2015),
 arXiv:1409.6094.

\bibitem{Haldane1988} 
 F.D.M. Haldane,
 Model for a quantum Hall effect without Landau levels: Condensed-matter realization of the "Parity Anomaly",
Phys. Rev. Lett. {\bf 61}, 2015--2018 (1988).

\bibitem{Mackenzie2003}
A.P. Mackenzie and Y. Maeno,
The superconductivity of Sr$_2$RuO$_4$ and the physics of spin-triplet pairing,
 Rev. Mod. Phys. {\bf 75}, 657--712 (2003).

\bibitem{Volovik1990} 
 G.E. Volovik,
Half quantum  vortices in the B phase of superfluid $^3$He,
JETP Lett. {\bf 52 }, 358--363 (1990).

\bibitem{KlinkhamerVolovik2005} 
F.R. Klinkhamer and G.E. Volovik, 
Emergent CPT violation from the splitting of Fermi points, 
Int. J. Mod. Phys. A {\bf 20}, 2795--2812 (2005); 
hep-th/0403037.


 \bibitem{HeikkilaVolovik2011} 
T.T. Heikkil\"a and G.E. Volovik,
Dimensional crossover in topological matter: Evolution of the multiple Dirac point in the layered system to the flat band on the surface,
Pis'ma ZhETF {\bf 93}, 63--68 (2011); JETP Lett. {\bf 93}, 59--65 (2011);
arXiv:1011.4185.

 \bibitem{HeikkilaVolovik2015} 
T.T. Heikkil\"a and G.E. Volovik,
Flat bands as a route to high-temperature superconductivity in graphite,
 arXiv:1504.05824.

\bibitem{Mikitik2006} 
G.P. Mikitik and Yu.V. Sharlai, 
Band-contact lines in the electron energy spectrum of graphite,
Phys. Rev. B {\bf 73}, 235112 (2006).

\bibitem{Mikitik2008} 
G.P. Mikitik and Yu.V. Sharlai,
The Berry phase in graphene and graphite multilayers,
Low Temp. Phys. {\bf 34}, 794--780 (2008).


\bibitem{Yudin2014}
D. Yudin, D. Hirschmeier, H. Hafermann, O. Eriksson, A.I. Lichtenstein  and M.I. Katsnelson,
Fermi condensation near van Hove singularities within the Hubbard model on the triangular lattice,
Phys. Rev. Lett. {\bf 112}, 070403 (2014).

\bibitem{Dolgopolov2014}
A.A. Shashkin, V.T. Dolgopolov, J.W. Clark, V.R. Shaginyan, M.V. Zverev and V.A. Khodel,
Merging of Landau levels in a strongly-interacting two-dimensional electron system in silicon,
 Phys. Rev. Lett. {\bf 112}, 186402 (2014).

\bibitem{KopninHeikkilaVolovik2011} 
N.B. Kopnin, T.T. Heikkil\"a and G.E. Volovik,
High-temperature surface superconductivity in topological flat-band systems,
Phys. Rev. B {\bf 83}, 220503(R) (2011);
arXiv:1103.2033.

\bibitem{Volovik1999}
G.E. Volovik, 
Fermion zero modes on vortices in  chiral superconductors,
JETP Lett. {\bf 70}, 609--614 (1999); cond-mat/9909426.


\bibitem{ReadGreen2000}
N. Read and D. Green,
Paired states of fermions in two dimensions with breaking of parity and time-reversal symmetries and the fractional quantum Hall effect,
Phys. Rev. B {\bf 61}, 10 267--10 297 (2000).


\bibitem{Ryu2002}
S. Ryu and  Y. Hatsugai, 
Topological origin of zero-energy edge states in particle-hole symmetric systems,
 Phys. Rev. Lett. {\bf 89}, 077002 (2002).

\bibitem{SchnyderRyu2011}
 A.P.  Schnyder and S. Ryu, 
Topological phases and flat surface bands in superconductors without inversion symmetry,
Phys. Rev. B {\bf 84}, 060504(R) (2011).

\bibitem{HeikkilaKopninVolovik2011} 
T.T. Heikkil\"a, N.B. Kopnin and G.E. Volovik,
Flat bands in topological media, 
JETP Lett. {\bf 94}, 233--239 (2011);
 arXiv:1012.0905.
 
 \bibitem{SilaevVolovik2014} 
M.A. Silaev, G.E.Volovik,
Andreev-Majorana bound states in superfluids,
JETP {\bf 119},   1042--1057 (2014),

\bibitem{Kane2015} 
Youngkuk Kim, B. J. Wieder, C. L. Kane, A. M. Rappe,
Dirac line nodes in inversion symmetric crystals,
arXiv:1504.03807 

\bibitem{EsquinaziHeikkila2014}
P. Esquinazi, T.T.  Heikkil\"a, Y.V. Lysogorskiy, D.A. Tayurskii, G.E. Volovik,
On the superconductivity of graphite interfaces,
Pis'ma ZhETF {\bf 100},  374--378 (2014); 
  arXiv:1407.2060.

\bibitem{TangFu2014}
E. Tang and L. Fu, 
Strain-induced partially flat band, helical snake states, and interface superconductivity in topological crystalline insulators,
Nature Phys. {\bf 10}, 964--969 (2014);
arXiv:1403.7523.


\bibitem{Ballestar HeikkilaEsquinazi2014}
A. Ballestar, T.T. Heikkil\"a, P. Esquinazi,
Size dependence of the Josephson critical behavior in pyrolytic graphite TEM lamellae,
 Supercond. Sci. Technol. {\bf 27}, 115014 (2014).


\end{thebibliography}
\end{document}